\documentclass[
  twocolumn,
  superscriptaddress,
  amsmath,amssymb,
  floatfix,
 aps, physrev,
]{revtex4-2}

\usepackage[unicode,colorlinks=true,allcolors=blue]{hyperref} % colored references and citations
\usepackage{graphicx}% Include figure files
\usepackage{dcolumn}% Align table columns on decimal point
\usepackage{bm}% bold math
\usepackage[separate-uncertainty=true]{siunitx}
\usepackage{here}
\usepackage{amssymb}
\usepackage[dvipsnames]{xcolor}
\usepackage{color}

\DeclareSIUnit\ecm{\text{\textit{e}}cm}

\definecolor{mplblue}{HTML}{1f77b4}
\definecolor{mplorange}{HTML}{ff7f0e}
\definecolor{mplgreen}{HTML}{2ca02c}

\begin{document}

\title{\textbf{Initial results of the TRIUMF ultracold advanced
    neutron source}}

\author{B.~Algohi}
\affiliation{University of Manitoba, Winnipeg, MB, Canada}
\author{D.~Anthony} % ugrad 2024
\affiliation{TRIUMF, Vancouver, BC, Canada}
\author{L.~Barr\'on-Palos}
\affiliation{Instituto de F\'isica, Universidad Nacional Aut\'onoma de M\'exico, Mexico City, Mexico}
\author{M.~Boss\'e} % ugrad 2025
\affiliation{TRIUMF, Vancouver, BC, Canada}
\author{M.P.~Bradley}
\affiliation{University of Saskatchewan, Saskatoon, SK, Canada}
\author{A.~Brossard}
\affiliation{TRIUMF, Vancouver, BC, Canada}
\author{T.~Bui} % ugrad 2024
\affiliation{University of Manitoba, Winnipeg, MB, Canada}
\author{J.~Chak}
\affiliation{TRIUMF, Vancouver, BC, Canada}
\author{R.~Chiba}
\affiliation{Simon Fraser University, Burnaby, BC, Canada}
\author{C.~Davis}
\affiliation{TRIUMF, Vancouver, BC, Canada}
\author{R.~de Vries} % ugrad 2024-2025
\affiliation{The University of Winnipeg, Winnipeg, MB, Canada}
\author{K.~Drury} % ugrad 2024-2025
\affiliation{TRIUMF, Vancouver, BC, Canada}
\author{B.~Franke}
\affiliation{TRIUMF, Vancouver, BC, Canada}
\affiliation{The University of British Columbia, Vancouver, BC, Canada}
\author{D.~Fujimoto}
\affiliation{TRIUMF, Vancouver, BC, Canada}
\author{R.~Fujitani}
\affiliation{Department of Nuclear Engineering, Kyoto University, Kyoto, Japan}
\affiliation{Institute for Integrated Radiation and Nuclear Science (KURNS), Kyoto University, Osaka, Japan}
\author{M.~Gericke}
\affiliation{University of Manitoba, Winnipeg, MB, Canada}
\author{D.~Georgescu}
\affiliation{TRIUMF, Vancouver, BC, Canada}
\author{P.~Giampa}
\affiliation{TRIUMF, Vancouver, BC, Canada}
\author{C.~Gibson}
\affiliation{TRIUMF, Vancouver, BC, Canada}
\author{R.~Golub}
\affiliation{North Carolina State University, Raleigh, NC, USA}
\author{K.~Hatanaka}\thanks{deceased}
\affiliation{Research Center for Nuclear Physics (RCNP), The University of Osaka, Osaka, Japan}
\author{T.~Hepworth} % ugrad 2023-2025
\affiliation{The University of Winnipeg, Winnipeg, MB, Canada}
\author{T.~Higuchi}
\affiliation{Institute for Integrated Radiation and Nuclear Science (KURNS), Kyoto University, Osaka, Japan}
\affiliation{Research Center for Nuclear Physics (RCNP), The University of Osaka, Osaka, Japan}
\author{G.~Ichikawa}
\affiliation{High Energy Accelerator Research Organization (KEK), Tsukuba, Ibaraki, Japan}
\author{I.~Ide}
\affiliation{Nagoya University, Nagoya, Aichi, Japan}
\author{S.~Imajo}\thanks{current affiliation:   RIKEN Center for Advanced Photonics, Wako, Saitama, Japan}
\affiliation{Research Center for Nuclear Physics (RCNP), The University of Osaka, Osaka, Japan}
\author{A.~Jaison}
\affiliation{University of Manitoba, Winnipeg, MB, Canada}
\author{B.~Jamieson}
\affiliation{The University of Winnipeg, Winnipeg, MB, Canada}
\author{K. Jorgensen-Fullam}
\affiliation{TRIUMF, Vancouver, BC, Canada}
\author{M.~Katotoka} % ugrad 2024-2025
\affiliation{The University of Winnipeg, Winnipeg, MB, Canada}
\author{S.~Kawasaki}
\affiliation{High Energy Accelerator Research Organization (KEK), Tsukuba, Ibaraki, Japan}
\affiliation{The Graduate University for Advanced Studies (Sokendai), Tsukuba, Ibaraki, Japan}
\author{M.~Kitaguchi}
\affiliation{Nagoya University, Nagoya, Aichi, Japan}
\author{W.~Klassen}
\affiliation{The University of British Columbia, Vancouver, BC, Canada}
\author{E.~Korkmaz}
\affiliation{The University of Northern BC, Prince George, BC, Canada}
\author{E.~Korobkina}
\affiliation{North Carolina State University, Raleigh, NC, USA}
\author{F.~Kuchler}\thanks{current affiliation:  Technical University of Munich, Munich, Germany}
\affiliation{TRIUMF, Vancouver, BC, Canada}
\author{C.~Lamb}
\affiliation{TRIUMF, Vancouver, BC, Canada}
\author{M.~Lavvaf}
\affiliation{University of Manitoba, Winnipeg, MB, Canada}
\author{T.~Lightbody}
\affiliation{TRIUMF, Vancouver, BC, Canada}
\author{T.~Lindner}
\affiliation{TRIUMF, Vancouver, BC, Canada}
\affiliation{The University of Winnipeg, Winnipeg, MB, Canada}
\author{N.~Lo} % ugrad fall 2024
\affiliation{TRIUMF, Vancouver, BC, Canada}
\author{S.~Longo}
\affiliation{University of Manitoba, Winnipeg, MB, Canada}
\author{K.W.~Madison}
\affiliation{The University of British Columbia, Vancouver, BC, Canada}
\author{Y.~Makida}
\affiliation{High Energy Accelerator Research Organization (KEK), Tsukuba, Ibaraki, Japan}
\affiliation{The Graduate University for Advanced Studies (Sokendai), Tsukuba, Ibaraki, Japan}
\author{J.~Malcolm} % ugrad fall 2024, winter 2025
\affiliation{TRIUMF, Vancouver, BC, Canada}
\author{J.~Mammei}
\affiliation{University of Manitoba, Winnipeg, MB, Canada}
\author{R.~Mammei}
\affiliation{The University of Winnipeg, Winnipeg, MB, Canada}
\author{Z.~Mao}
\affiliation{The University of British Columbia, Vancouver, BC, Canada}
\author{C.~Marshall}
\affiliation{TRIUMF, Vancouver, BC, Canada}
\author{J.W.~Martin}
\affiliation{The University of Winnipeg, Winnipeg, MB, Canada}
\author{R.~Matsumiya}
\affiliation{TRIUMF, Vancouver, BC, Canada}
\affiliation{Research Center for Nuclear Physics (RCNP), The University of Osaka, Osaka, Japan}
\author{M.~McCrea}
\affiliation{The University of Winnipeg, Winnipeg, MB, Canada}
\author{E.~Miller}
\affiliation{The University of British Columbia, Vancouver, BC, Canada}
\author{M.~Miller}
\affiliation{McGill University, Montreal, QC, Canada}
\author{K.~Mishima}
\affiliation{Research Center for Nuclear Physics (RCNP), The University of Osaka, Osaka, Japan}
\affiliation{Nagoya University, Nagoya, Aichi, Japan}
\affiliation{High Energy Accelerator Research Organization (KEK), Tsukuba, Ibaraki, Japan}
\author{T.~Mohammadi}
\affiliation{University of Manitoba, Winnipeg, MB, Canada}
\author{T.~Momose}
\affiliation{The University of British Columbia, Vancouver, BC, Canada}
\affiliation{TRIUMF, Vancouver, BC, Canada}
\author{M.~Nalbandian} % ugrad 2025
\affiliation{The University of British Columbia, Vancouver, BC, Canada}
\author{T.~Okamura}
\affiliation{High Energy Accelerator Research Organization (KEK), Tsukuba, Ibaraki, Japan}
\affiliation{The Graduate University for Advanced Studies (Sokendai), Tsukuba, Ibaraki, Japan}
\author{S.~Pankratz} % ugrad 2025
\affiliation{The University of Winnipeg, Winnipeg, MB, Canada}
\author{R.~Patni} % ugrad fall 2024, winter 2025
\affiliation{TRIUMF, Vancouver, BC, Canada}
\author{R.~Picker}
\affiliation{TRIUMF, Vancouver, BC, Canada}
\affiliation{Simon Fraser University, Burnaby, BC, Canada}
\author{V.~Purcell}
\affiliation{TRIUMF, Vancouver, BC, Canada}
\author{K.~Qiao}
\affiliation{Research Center for Nuclear Physics (RCNP), The University of Osaka, Osaka, Japan}
\affiliation{Graduate School of Science, The University of Osaka, Osaka, Japan}
\author{W.D.~Ramsay}
\affiliation{TRIUMF, Vancouver, BC, Canada}
\author{W.~Rathnakela}
\affiliation{University of Manitoba, Winnipeg, MB, Canada}
\author{T.~Reimer} % ugrad 2025
\affiliation{The University of Winnipeg, Winnipeg, MB, Canada}
\author{D.~Salazar}
\affiliation{Simon Fraser University, Burnaby, BC, Canada}
\author{J.~Sato}
\affiliation{Nagoya University, Nagoya, Aichi, Japan}
\author{W.~Schreyer}
\affiliation{TRIUMF, Vancouver, BC, Canada}
\affiliation{Physics Division, Oak Ridge National Laboratory, Oak Ridge, TN, USA}
\author{T.~Shima}
\affiliation{Research Center for Nuclear Physics (RCNP), The University of Osaka, Osaka, Japan}
\author{H.M.~Shimizu}
\affiliation{Nagoya University, Nagoya, Aichi, Japan}
\author{S. Siddiqui}
\affiliation{TRIUMF, Vancouver, BC, Canada}
\author{S.~Sidhu}
\affiliation{TRIUMF, Vancouver, BC, Canada}
\author{S.~Stargardter}
\affiliation{University of Manitoba, Winnipeg, MB, Canada}
\author{R.~Stutters} % ugrad 2025
\affiliation{TRIUMF, Vancouver, BC, Canada}
\author{P.~Switzer} % ugrad summer 2024
\affiliation{The University of Winnipeg, Winnipeg, MB, Canada}
\author{I.~Tanihata}
\affiliation{Research Center for Nuclear Physics (RCNP), The University of Osaka, Osaka, Japan}
\author{Tushar}
\affiliation{University of Manitoba, Winnipeg, MB, Canada}
\author{M. Uzair}
\affiliation{TRIUMF, Vancouver, BC, Canada}
\author{S.~Vanbergen}
\affiliation{The University of British Columbia, Vancouver, BC, Canada}
\affiliation{TRIUMF, Vancouver, BC, Canada}
\author{W.T.H.~van~Oers}
\affiliation{TRIUMF, Vancouver, BC, Canada}
\author{N.~Yazdandoost}
\affiliation{TRIUMF, Vancouver, BC, Canada}
\author{Q.~Ye}
\affiliation{The University of British Columbia, Vancouver, BC, Canada}
\author{A.~Zahra}
\affiliation{University of Manitoba, Winnipeg, MB, Canada}
\author{M.~Zhao} % ugrad winter-summer 2024
\affiliation{TRIUMF, Vancouver, BC, Canada}

\collaboration{TUCAN Collaboration}

\date{\today}

\begin{abstract}
We report the first results on ultracold neutron production from a new
spallation-driven superfluid $^4$He (He-II) source at TRIUMF, which is
being prepared for a new, precise measurement of the neutron electric
dipole moment.  A total of $(9.3 \pm 0.8)\times 10^{5}$ ultracold
neutrons were observed at a proton beam current of \SI{37}{\uA}, when
the target was irradiated for a period of \SI{60}{\s}.  The results
are in fair agreement with expectations based on a detailed simulation
of neutron transport and ultracold neutron source cryogenics.  There
is some indication that the new source might not be as limited by the
conduction of heat through the He-II as originally expected.  The
results indicate that the source is likely to make its ultimate
production goals, once the liquid deuterium cold moderator system is
completed, with the expectation that $5.7\times 10^7$~UCNs would be
detected in the same experiment with full liquid levels.  This would,
for example, correspond to delivery of $1.4\times 10^6$~UCNs delivered
to each of two nEDM measurement cells, and a statistical uncertainty
of $1\times 10^{-27}~e$cm on the neutron EDM in 280 days of running.
\end{abstract}

\maketitle

%\section{\label{sec:intro}Introduction and Motivation}

The neutron electric dipole moment (nEDM) is an experimental
observable of high importance in fundamental physics because it
violates time-reversal symmetry and therefore CP (charge-parity)
symmetry~\cite{bib:pospelov,bib:engel,bib:chuppall}, the symmetry
relating the interactions of particles to those of their antiparticle
counterparts. To date, all experiments have found the nEDM to be
consistent with zero.  Improving the experimental precision places
tighter constraints on new sources of CP violation beyond the Standard
Model.  Conversely, if a small but non-zero nEDM were measured, it
would herald a discovery of new physics.  Even if ascribed to the
CP-violating $\bar{\theta}$ parameter of the strong sector, the
mystery of a small but non-zero $\bar{\theta}$ would create a new
problem for the Standard Model.

Recent theoretical work addressing the physics impact of an even more
precise measurement of the nEDM has focused on three general (and
overlapping) themes: (1) new sources of CP violation beyond the
Standard Model~\cite{bib:cirigliano,bib:crivellin}, (2) baryogenesis
scenarios, especially new physics contributions to electroweak
baryogenesis inspired scenarios~\cite{bib:bell,bib:hou} and (3) the
strong CP problem, related to searches for
axions and axionless solutions~\cite{bib:carena,bib:mimura,bib:peinado,bib:psiaxion}.  Because
of these connections, better measurements of the nEDM are of vital
importance in particle physics and early universe cosmology.

A recent measurement performed using ultracold neutrons (UCNs) at the
Paul Scherrer Institute (PSI, Villigen, Switzerland) determined an
upper bound on the nEDM,
$|d_{\mathrm{n}}|<\SI{1.8e-26}{\ecm}$~(\SI{90}{\percent}
C.L.)~\cite{bib:psi2020}.  Next generation UCN EDM experiments are in
preparation at a variety of sites worldwide, and are aiming to improve
the result by an order of magnitude or more.  Experiments are planned
at Institut Laue-Langevin (ILL, Grenoble, France)~\cite{bib:panedm},
PSI~\cite{bib:n2edm}, and Los Alamos National Laboratory (LANL, Los
Alamos, NM, USA)~\cite{bib:lanledm}, in addition to our effort at
TRIUMF (Canada's particle accelerator centre, Vancouver, BC,
Canada)~\cite{bib:npn}.

It is presently unclear which of the ongoing experiments will
ultimately yield the best sensitivity.  While all these experiments
aim to use UCNs stored in measurement chambers to perform the nEDM
experiment, they differ in the UCN source technology employed.  The
LANL and PSI projects use solid ortho-deuterium (sD$_2$) sources
driven by spallation~\cite{bib:lanlsource,bib:psisource}.  The
SuperSUN project uses a He-II source placed within a cold neutron
beamline at the ILL reactor and has recently demonstrated UCN
production consistent with expectations~\cite{bib:supersun}.  The
TRIUMF UltraCold Advanced Neutron (TUCAN) source uses a He-II source
also, but couples it to a spallation target using heavy water (D$_2$O)
and liquid deuterium (LD$_2$) neutron moderators, with the potential
to increase the flux of cold neutrons entering the UCN production
volume.

Although the production cross-section of UCNs is larger in
sD$_2$~\cite{bib:yumalikgolub,bib:atchison,bib:frei} than for
superfluid $^4$He (He-II)~\cite{bib:golub77}, due to the availability
of more excitation modes, the losses of UCN by upscattering (from both
ortho- and para-deuterium), and hydrogen
contamination~\cite{bib:lanlucn,bib:chenyu} limit the lifetime of
neutrons in the sD$_2$ to tens of milliseconds, and losses can be
worsened by surface frost~\cite{bib:psincsulanl}.  In He-II, the
losses can be significantly lower, and are limited by phonon
upscattering.  It has been determined experimentally that the
upscattering losses are best described by a two phonon process, giving
a loss rate proportional to $T^7$, where $T$ is the temperature of the
He-II~\cite{bib:zimmer,bib:leung,bib:yoshiki}.  If the temperature of
the He-II can be reduced below \SI{1}{\K}, the neutron storage
lifetime within the He-II can be greater than
\SI{100}{\s}~\cite{bib:yoshiki}.

The TUCAN source uses the \SI{483}{\MeV} proton beam from the TRIUMF
cyclotron.  The basis of the TUCAN approach involves a
spallation-driven, superfluid $^4$He (He-II) UCN source connected to a
room-temperature nEDM experiment.  In this Letter, we report the first
results on UCN production from the TUCAN source.

%\section{\label{sec:source}UCN source and performance expectations}
{\it UCN source and performance expectations}---Our UCN source is
based on previous work on a prototype vertical UCN source reported in
Refs.~\cite{bib:masuda1,bib:masuda2}.  Originally operated at RCNP
Osaka, this source was installed at TRIUMF in 2017, at a new proton
beamline fed by a fast kicker magnet, capable of
delivering~\SI{40}{\uA} to a new tungsten spallation
target~\cite{bib:ucnbeamline,bib:kicker}.  The system was used for
experiments on UCN production~\cite{bib:physrevc}, transport,
storage~\cite{bib:tailsection}, polarization, and
detection~\cite{bib:shrthesis}.  In 2020--2021, the vertical source
was decommissioned to make way for a new, significantly upgraded UCN
source.  The source is now complete to a degree that first results on
UCN production can be reported.

\begin{figure} 
%trim=left bottom right top
\centering \includegraphics[width=\columnwidth, page=1,trim=0cm 2.35cm 0cm 2.4cm,clip]{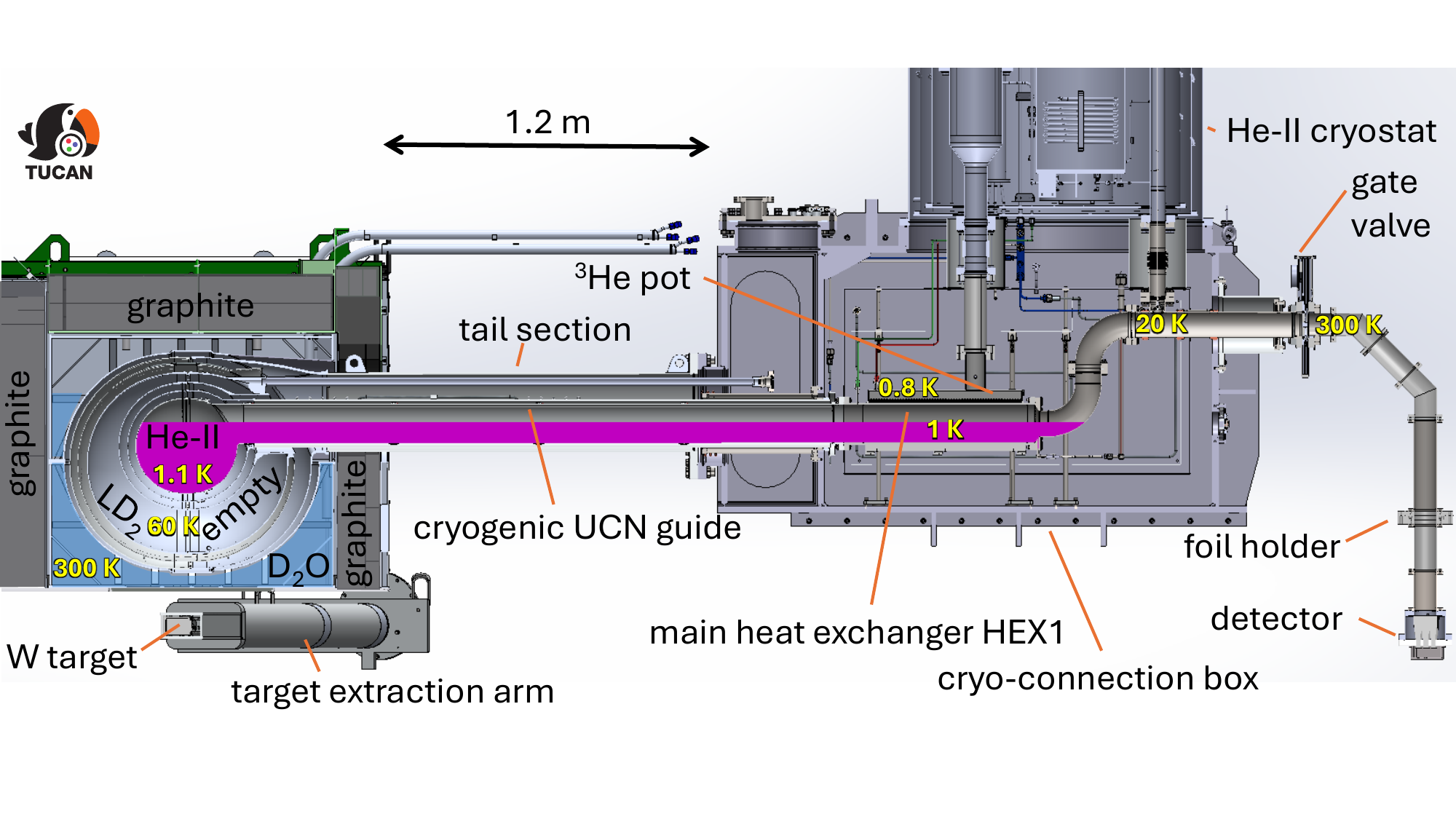}
\caption{The UCN source and detector configuration used in these
  experiments.  The pink volume indicates the fill level of the He-II
  for these experiments, the light blue the D$_2$O level.  The proton beam impinges upon the W target horizontally, approximately perpendicular to the tail section orientation (out of the page).\label{fig:detector-cad}}
\end{figure}

The new UCN source features a \SI{27}{\litre} UCN production volume
(the bulb labeled He-II in Fig.~\ref{fig:detector-cad}) which is
significantly larger than the \SI{8}{\litre} volume used in the
vertical source.  Because the UCNs exit the source horizontally, the
new source is referred to as the horizontal source.  The vertical
source could reliably handle 300~mW of heat load to the He-II whereas
the horizontal source is optimized for heat loads up to
\SI{10}{\W}~\cite{bib:shinsuke,bib:okamura}.  A new $^3$He
refrigerator and a larger capacity helium pumping system enables the
additional cooling power.  Additionally, a new large-area
$^3$He--$^4$He heat exchanger (HEX1 in Fig.~\ref{fig:detector-cad})
was built to be compatible with both UCN transport and heat transfer
requirements, resolving a severe limitation of the vertical
source~\cite{bib:shrthesis}.  An optimized moderator
system~\cite{bib:moderators} featuring an LD$_2$ cold moderator will
be used to produce a large cold neutron flux, which is expected to
provide another significant improvement.
 
% Request end matter

%\begin{figure}
%\centering
%\includegraphics[width=\columnwidth]{overhead-view-april-2024_cropped.pdf}
%\caption{Overhead view of the UCN source facility (April 2024).  Lines
%  display the underlying proton beam path (red) and sketch the
%  existing and planned UCN guide paths (blue).\label{fig:overhead}}
%\end{figure}

%An overhead view of the facility is shown in Fig.~\ref{fig:overhead}.
%The photograph was taken in April 2024, just before the UCN source was
%covered in shielding blocks.  The facility is now complete to the
%The UCN source has been operated in a month-long cryogenic
%test run, and the magnetically shielded room (MSR) is routinely in use
%for magnetometer and coil testing, in preparation for UCN runs planned
%for late 2025.

Detailed estimates have been conducted for UCN production and
extraction~\cite{bib:moderators,bib:ss,bib:ssthesis} and we now
present a survey of the expected results once the UCN source has been
fully completed.  Production estimates were based on a Monte Carlo
N-Particle (MCNP) model of the UCN source, which included the target
and moderators, and cold neutron fluxes were converted to UCN
production based on~\cite{bib:korobk,bib:schmidt-wellenburg}. UCN
transport simulations using PENTrack~\cite{bib:pentrack} calculated
losses within the He-II and in transport to the EDM experiment.  The
losses in the He-II due to multiphonon processes were estimated based
on a 1D thermal model of the He-II volume, assuming that heat
conduction was in the quantum-turbulent Gorter-Mellink regime of heat
conduction~\cite{bib:vansciver}.  Based on the simulations, when
driven by a \SI{40}{\uA} proton beam, the source is expected to
produce \SI{1.4e7}{UCNs/\s}, with beam heating of \SI{8.1}{\W} to the
He-II and its containment vessel at \SI{1.1}{\K}. This is more than
two orders of magnitude larger than the UCN production rate of the
vertical source.  Using a 125~s period of target irradiation, an
estimated total of \SI{1.38e7}{UCNs} would be loaded into the EDM
measurement cells prior to initiating the Ramsey (frequency
measurement) cycle.  Using reasonable values for lifetimes and
spin-coherence times of the UCNs, this corresponds to a statistical
uncertainty of the nEDM of $\sigma(d_{\mathrm{n}})=\SI{3e-25}{\ecm}$
per cycle.  Using conservative assumptions for the running time
available per day, a statistical uncertainty of
$\sigma(d_{\mathrm{n}})=\SI{e-27}{\ecm}$ would be achieved within 280
measurement days (16 hours of uptime per day)~\cite{bib:ss}.

%\section{\label{sec:apparatus}First UCN experiments}

{\it First UCN experiments}---First UCN experiments were conducted
with a detector inside the radiation shielding, directly at the exit
of the UCN source (Fig.~\ref{fig:detector-cad}).  UCN guides were
connected to divert the UCNs downward into the detector. A
\SI{100}{\um} Al foil could be installed in the foil holder indicated
in Fig.~\ref{fig:detector-cad}.  The detector was located
\SI{1.11}{\m} below the UCN guide exit from the source, and
\SI{52}{\cm} below the UCN production volume. The UCN source was
filled with isotopically pure superfluid $^4$He up to a fill level of
\SI{27}{\cm} out of a total diameter of \SI{36}{\cm} of the UCN
production volume (bulb).  The isotopically pure $^4$He was purchased
from Lancaster (UK), with a specified $^3$He content $<10^{-11}$.
Measurements performed by the vendor set an upper bound of $<6\times
10^{-12}$, implying a lifetime due to captures of $>6000$~s.  This is
not expected to cause significant UCN losses.  The fill level
was limited by the amount of isotopically pure $^4$He on hand at
TRIUMF.

A typical experimental cycle would consist of irradiating the
spallation target with the proton beam, waiting a set period of time,
and then opening a UCN-compatible gate valve to count the number of
UCNs.  By adjusting the irradiation time, or the
waiting time with valve closed after irradiation, the storage lifetime
of the UCNs in the source can be deduced.

Results presented in this paper were from a lithium-loaded
scintillating glass detector with a detection efficiency of
90\%~\cite{bib:blair}.  A second detector, a Strelkov DUNia-10
proportional counter based on $^3$He, was used for systematic checks.

%\section{\label{sec:cryo}Cryogenic performance}
{\it Cryogenic performance}---The cryogenic performance of the source
was monitored during beam testing, and was further tested in offline
runs using heaters.  The base temperature was found to be \SI{0.9}{K} with a resting heat load of $\approx$\SI{2}{W}.  The slope of $^3$He cooling power (measured by its pumping rate) with beam current was consistent with expectations based on our MCNP simulations, rising to \SI{4}{W} at \SI{10}{\uA} current.

\begin{figure}
\begin{center}
  \includegraphics[width=\columnwidth]{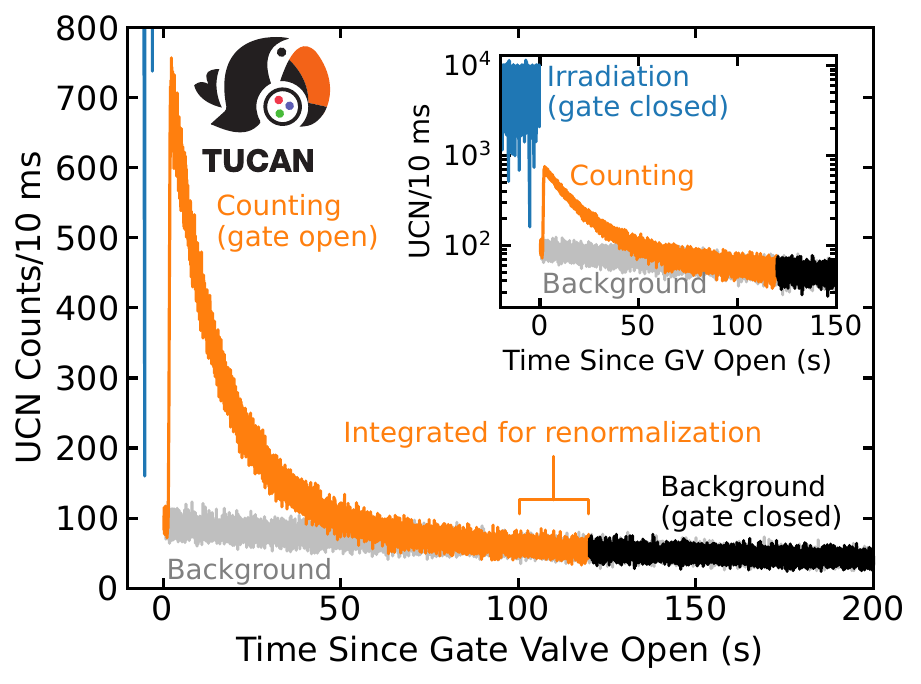}
\caption{UCN counts during a measurement cycle. The source was irradiated at \SI{37}{\uA} for \SI{60}{\s} (blue), after which the UCN gate valve was opened for \SI{120}{\s} (orange$\rightarrow$black). After subtracting the background run (grey), the integrated UCN counts during the valve-open period were $(9.3\pm 0.8)\times 10^5$. The inset shows the same data in a semi-log plot.}
\label{fig:bigpulse}
\end{center}
\end{figure}

{\it UCN production results}---The source was characterized in UCN
production runs conducted in June and August 2025.  We now describe
the conditions and measurement cycle which resulted in the data
presented in Fig.~\ref{fig:bigpulse}.  The current in Beamline 1A
\cite{bib:ucnbeamline} was set to \SI{110}{\uA}.  The kicker
\cite{bib:kicker}, which diverts the beam into Beamline 1U (toward the UCN
source), was set to select 1 out of 3 pulses when switched on, so that
the time-averaged beam current delivered to the spallation target was
\SI{36.7}{\uA}.  The kicker was switched on for a \SI{60}{\s}
irradiation time.  During the irradiation time, many background
neutrons were observed in the $^6$Li detector.  At the end of the
irradiation, the UCN gate valve was opened without delay, so that UCNs
from the source could be detected (the orange region in
Fig.~\ref{fig:bigpulse}).  After \SI{120}{\s} of counting time, the
gate valve was closed to monitor the background after the measurement
period.  Measurement cycles generally alternated with background
measurement cycles where the same target irradiation occurred but the
gate valve was not opened.  Detector backgrounds were reduced by
applying cuts to the waveform parameters of each detected event
(PSD~$>0.3$ and $Q_{\mathrm{L}}>2000$, as defined
in~\cite{bib:blair}).

For the run shown in Fig.~\ref{fig:bigpulse}, the total number of UCNs
counted was $(9.3\pm 0.8)\times 10^5$.  The uncertainty is dominated
by background subtraction, which will be discussed momentarily.  The
measurement cycle was repeated for various beam currents, up to the
maximum possible by our current setup, \SI{36.7}{\uA}.  In future
work, we will increase the current slightly to \SI{40}{\uA}
\cite{bib:kicker}.

\begin{figure}
\begin{center}
\includegraphics[width=\columnwidth]{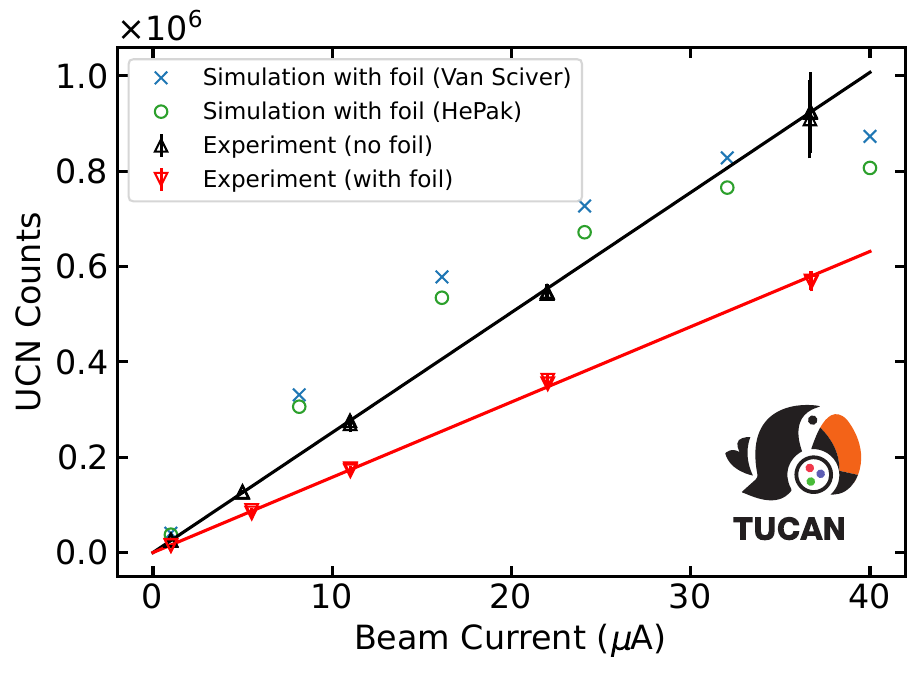}
\caption{Integrated UCN counts in the \SI{120}{\s} counting period as
  a function of the average beam current during the preceding
  \SI{60}{\s} irradiation period. Most of the error bars are similar
  to the vertical size of the data points, save the measurements at
  the highest current with no foil (see text).  The data are fitted to
  a straight line with forced zero intercept. Data are compared with
  simulations (which all assume the presence of a foil, see text).}
\label{fig:ucn-vs-current}
\end{center}
\end{figure}

Keeping the beam-on period at \SI{60}{\s}, the current impinging on
the target was adjusted by changing the kicker duty cycle.  The number
of UCN counts was measured as a function of the beam current
(Fig.~\ref{fig:ucn-vs-current}).  The data were fitted to a straight
line with forced zero intercept, giving a slope of \SI{2.52\pm0.02
  e4}{UCNs/\uA} with reduced $\chi^2_\nu$ of 0.32 on $\nu=14$ degrees
of freedom.

Fig.~\ref{fig:ucn-vs-current} also shows measurements with a vacuum
separation Al foil of thickness \SI{100}{\um} between the detector and
storage volume.  Its presence results in a reduction to
\SI{63\pm1}{\percent} of the UCN counts without the foil.  The slope
of the linear fit is \SI{1.57\pm 0.02 e4}{UCNs/\uA} with a reduced
$\chi^2_\nu$ of 0.44 ($\nu=10$).  The measurements with foil were
taken about a month after those without.  Simulations (described in
the next section) were conducted including the
foil~\cite{bib:pentrack}.  Therefore, the most accurate comparisons of
data to simulation are to compare the red data points
(\textcolor{red}{$\blacktriangledown$}) to either the blue simulated
points (\textcolor{mplblue}{$\times$}) or green simulated points
(\textcolor{mplgreen}{$\circ$}) in Fig.~\ref{fig:ucn-vs-current}.  The
comparison will be discussed further in the next section, where the
assumptions of the simulated points are described.

In the future, the Al foil will be housed within a superconducting
polarizer magnet.  The strong field within the magnet will accelerate
high-field-seeking UCNs through the foil, reducing UCN losses.  The
ultimate goal of the foil is to keep the He-II production volume free
of cryopumped contaminants.

The main systematic uncertainties in the number of UCN counts came
from background subtraction and UCN detector deadtime.  The background
arose mostly from $\gamma$ rays emitted from activated components near
the detector, which for these experiments was located in the harsh
environment inside the radiation shielding.  In the \SI{120}{\s}
counting period, background accounted for \SI{47}{\percent} of the
measured counts, in the highest current runs, and a valve-closed run
was used to subtract the background.  Backgrounds tended to depend on
the history of the irradiations leading up to each measurement.  To
account for this, the integrated background was renormalized by the
ratio of the counts during the last \SI{20}{\s} of the \SI{120}{\s}
data window of the background and measurement runs
(Fig.~\ref{fig:bigpulse}).  This resulted in greatly reduced scatter
in the background-subtracted number of UCN counts for runs with open
gate valve.  The renormalization factor for the background varied
between 0.9 and 1.2.  The uncertainty attributed to the
renormalization was assigned an error of half the size of the maximum
deviation from unity, {\it i.e.}  \SI{10}{\percent}.  This resulted in
a systematic error on the counts of about 9\% for the highest current
points.  The error bars displayed in Fig.~\ref{fig:ucn-vs-current}
contain the statistical and systematic errors added in quadrature, but
they are dominated by this systematic error.  The largest error bar is
seen for the highest current data point with no foil.  There were
larger backgrounds for this data point because our beamline was tested
at high currents for several hours prior to this run.  This was
necessary to confirm the safety of our radiation shielding, which had
not been checked at such high currents before.  Other data points have
uncertainties similar in size as the symbols.  Multiple points are
also displayed at each beam current, which show good reproducibility.

Other uncertainties considered were due to deadtime and pileup. The
acquisition system used a CAEN V1725 module to read out the detectors
in a nearly continuous fashion.  Offline tests indicated that deadtime
corrections were \SI{<1}{\percent}.  Regarding pileup, the highest
singles rates for the $^6$Li detector were found to be \SI{40}{\kHz},
and the waveform digitizer time window was \SI{200}{\ns}.  Thus, random
coincidences within the time window between the singles rate were
\SI{<0.01}{\percent}.  These uncertainties are negligible compared
with the systematic error assigned to background subtraction, even at
smaller currents.

\begin{figure}
\begin{center}
\includegraphics[width=\columnwidth]{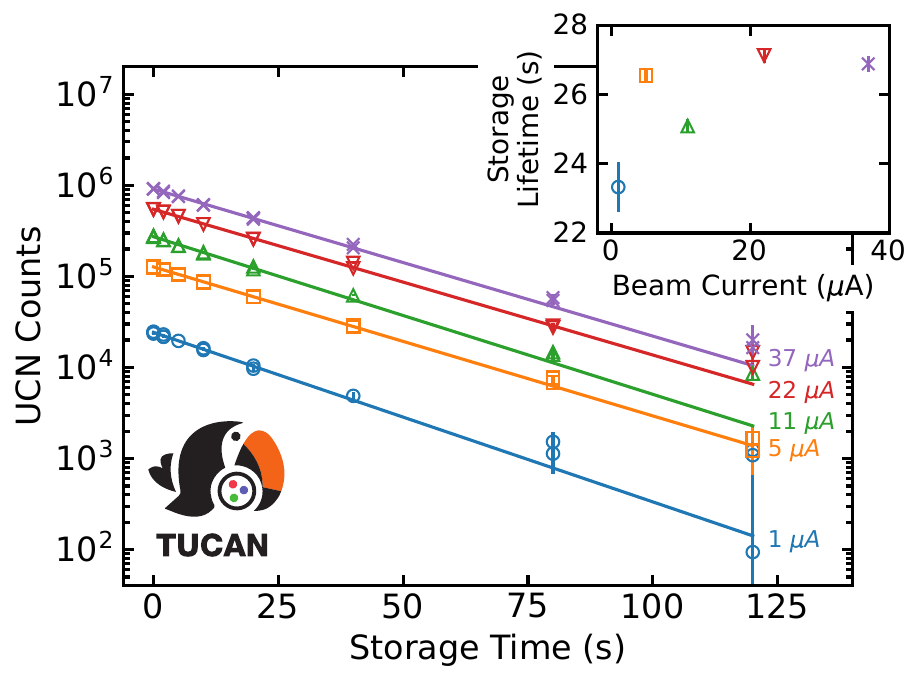}
\caption{Measurements of the UCN storage lifetime up to the UCN valve,
  as a function of the average beam current during the
  preceding~\SI{60}{\s} irradiation period.  Inset: summary of results
  for the measured lifetime from the exponential fits shown in the
  main graph.}
\label{fig:lifetime}
\end{center}
\end{figure}

We also conducted experiments to measure the storage lifetime of the
UCNs in the volume up to the UCN valve, by varying the valve opening
time after the beam was switched off.  An overview of these
measurements is presented in Fig.~\ref{fig:lifetime}.
We found the storage lifetime was in the range of 23--\SI{27}{\s}, and
that it did not strongly depend on current.  Our thermal model
(described below) would have predicted that the storage lifetime would
be smaller at higher current.  Based on the lack of
temperature-dependence, we conclude that wall losses seem to be
dominant for our operating parameters.

%\section{\label{sec:sims}Comparison to simulation and discussion}
{\it Comparison to simulation and discussion}---In
Fig.~\ref{fig:ucn-vs-current}, the UCN production data are compared
with expectations based on simulations of neutron transport and
cryogenic modeling of the UCN source.  To predict the UCN yield and
storage lifetime, the simulations were adapted from those discussed
and referenced earlier in this Letter.  Simulations assume a detector
efficiency of 90\%, consistent with~\cite{bib:blair}.  The MCNP
calculations to determine the UCN production rate reflected that the
heavy-water moderator was only partially filled (\SI{390}{\litre} out
of \SI{546}{\litre}, indicated schematically in
Fig.~\ref{fig:detector-cad}) and that the LD$_2$ moderator vessel was
empty.  PENTrack simulations of UCN transport assumed the UCN source
was fully filled to its design level with isotopically pure superfluid
$^4$He.  They used the formulation of Van Sciver~\cite{bib:vansciver}
or HEPAK~\cite{bib:hepak} to determine the temperature profile in the
UCN source based on heat conduction in turbulent He-II (the
Gorter-Mellink regime).  Since only constant material properties can
be simulated, the beam heat load during irradiation periods was chosen
as steady-state input to the calculations.

To set a scale, the simulations showed that when fully filled with
LD$_2$, D$_2$O, and He-II, and driven at \SI{40}{\uA} for a period of
\SI{60}{\s}, $5.7\times 10^7$~UCNs would be detected when the UCN gate
valve is opened immediately after the irradiation period.

When the MCNP simulations were repeated for the situation of the
LD$_2$ volume being empty, the He-II and D$_2$O levels being reduced,
and the beam current being \SI{36.7}{\uA}, the production rate dropped
from \SI{1.4e7}{UCNs/\s} to \SI{2.3e5}{UCNs/\s} (a factor of 61
reduction) and the beam-associated heat load to the He-II was reduced
from \SI{8.2}{\W} to \SI{4.8}{\W}.  The results of \cite{bib:ssthesis}
for \SI{60}{\s} irradiation time as a function of proton beam current,
were reduced by the factor of 61 (scaling with the UCN production
rate), and are shown by the green circles in
Fig.~\ref{fig:ucn-vs-current}.  The reduction in the UCN production
rate in this case is mostly the result of the absence of the LD$_2$
cold moderator.  The green circles in Fig.~\ref{fig:ucn-vs-current}
used the thermal conductivity function of~\cite{bib:hepak} whereas the
blue crosses use the thermal conductivity function
of~\cite{bib:vansciver}.

The more recent MCNP simulations were also updated to use a new
superfluid-helium scattering kernel from~\cite{bib:kkhleung}.  Our
previous work \cite{bib:moderators} had followed the suggestion
of~\cite{bib:albertpaper} in reducing the $^4$He density to account
for the difference.

The PENTrack and thermal distribution calculations were not repeated.
This is potentially problematic in that the heat conduction in the
Gorter-Mellink regime scales as heat flux cubed.  The reduction in the
liquid level in the long, horizontal channel doubles the heat flux
while the reduction in heat load due to lower fluid levels applies a
factor of 0.59 scaling by the beam heat load.  Thus, these effects
counteract one another and would result in a similar temperature
gradient in the horizontal section.  It should also be noted that the
influence of UCN losses in the superfluid ($\propto T^7$) would be
somewhat reduced, since there is less superfluid (replaced by helium
vapour) in the UCN source.  The scaling of the simulation comes with
these caveats.

In general, the data (the red
points:~\textcolor{red}{$\blacktriangledown$} in
Fig.~\ref{fig:ucn-vs-current}) lie somewhat below our predicted values
(blue~\textcolor{mplblue}{$\times$} or
green~\textcolor{mplgreen}{$\circ$}).  The difference tends to get
smaller at higher beam currents.  While the lower than expected output
might at first seem discouraging, predictions of the yields of UCN
sources are difficult.  We are encouraged by the level of agreement as
presented, because even rescaling the simulation to match the data
would result in world-leading performance once the LD$_2$ moderator
system is complete.

A prominent difference between the simulation and the experimental
data is that the experimental data rises linearly with beam current,
whereas the simulated UCN counts tend to saturate.  The saturation
observed in the simulation is caused by increasing temperature $T$ in
the UCN production volume and horizontal guide, which reduces the
neutron storage lifetime in the He-II as $1/\tau_{\rm He}=BT^7$ where
we assumed $B=\SI{0.016}{K^{-7}s^{-1}}$~\cite{bib:ssthesis}.  In our
thermal simulations of the UCN source, the temperature increase is not
uniform and is caused by higher heat load increasing (1) the
temperature gradient inside the superfluid in the long, horizontal
channel (via the Gorter-Mellink heat conductivity function); (2) the
temperature gap between the superfluid helium and our main heat
exchanger HEX1 (via Kapitza resistance); (3) the temperature gap
between the copper of HEX1 and the $^3$He (via extrapolated $^3$He
boiling data and Kapitza resistance); and (4) the temperature of the
$^3$He at the top of HEX1.
% Since temperature differences across
% interfaces and inside superfluid helium tend to decrease for higher
% temperatures, the overall temperature increase in the production
% volume is somewhat softened compared to what might be expected if HEX1
% were kept at constant temperature.

The data does not tend to saturate with beam current, being rather
well-described by the linear fit in Fig.~\ref{fig:ucn-vs-current}, in
disagreement with the simulations.  This could signify that future
liquid helium UCN sources similar to ours~({\it
  e.g.}~\cite{bib:albertpaper,bib:turlybekuly}) could be capable of
supporting higher power and higher neutron fluxes than anticipated. A
more detailed analysis of the heater and cryogenic data will be
necessary to determine whether the deviation from the model is due to
the transient nature of the heating, the nature of the heat conduction
(be it in the quantum-turbulent Gorter-Mellink regime or otherwise),
the assumptions for the UCN phonon upscattering loss parameter $B$, or
wall losses, or other errors attributable to the naive scaling of the
simulations.  The results could bear on quantum turbulence in He-II,
which has not been measured in this range of temperatures, channel
size, and heat flux.  The slope of the data with current is less steep
than the simulation, for smaller currents.  This may indicate that
wall losses, which should dominate at lower temperatures, are larger
than those assumed in the simulation.  The larger wall losses reduce
the sensitivity of the data to the predicted saturation effect.

The extrapolation of our results to the situation with full liquid
levels is rather straightforward.  None of the geometry will change,
save the level of the He-II, the level of the D$_2$O, and the presence
of LD$_2$.  Historically, the most uncertain extrapolation when trying
to scale up UCN source technology has been to do with temperature and
volume-dependent effects involving the UCN converter materials.  In
our new UCN source, we have demonstrated the ability to extract a
large number of UCN through the free surface of a He-II production
volume, and to do so reliably and repeatedly in runs that lasted a
week.  Higher heat loads to the He-II mimicking the situation with
full liquid levels have been studied by applying additional heat with
resistive heaters during the beam-on period, and while the results are
still being analyzed, they are encouraging.  The estimated factor of
61 improvement for full UCN source operations is in reach.

%\section{Conclusion and future prospects}
{\it Conclusion and future prospects}---In the highest current
\SI{36.7}{\uA} beam pulse, with \SI{60}{\s} irradiation time, we have
detected \SI{9.3\pm0.8e5}{UCNs} from the TUCAN source.  This is
projected to increase by a factor of 61, once our LD$_2$ cold neutron
moderator is installed and operating, and the heavy water moderator
tank and the UCN production volume are fully filled.  The factor is
based on our knowledge of fluid levels and on well-benchmarked MCNP
simulations.  The UCN losses are not expected to change significantly
from the experimental results that we present here.  The data bode
well for a significant improvement in UCN production from the TUCAN
source.

We do not see evidence of saturation of the UCN counts with proton
beam current, predicted by a thermal model involving extrapolation of
the Gorter-Mellink heat conductivity function to our temperature range
and channel dimension.  This potentially removes an expected
limitation to our source, which could open a new pathway to even more
intense sources of UCNs.

% We have acquired further data on UCN lifetimes in He-II, at a variety
% of temperature settings, and heat applied to different sections of the
% He-II system using resistive heaters.  We plan to report these results
% in a future publication.

%\section{Acknowledgments}

%\begin{acknowledgments}
{\it Acknowledgments}---We would like to sincerely thank the
Mechanical Engineering Center at KEK and the TRIUMF Design and
Fabrication, Cryo, Accelerator Systems and Operations groups.

We gratefully acknowledge the support of the Canada Foundation for
Innovation; the Canada Research Chairs program; the Natural Sciences
and Engineering Research Council of Canada (NSERC) SAPPJ-2016-00024,
SAPPJ-2019-00031, SAPPJ-2023-00029, and SAPPJ-2024-00030; British
Columbia Knowledge Development Fund; Research Manitoba; JSPS KAKENHI
(Grant Nos. 18H05230, 19K23442, 20KK0069, 20K14487, and 22H01236,
25H00652); JSPS Bilateral Program (Grant No. JSPSBP120239940); JST
FOREST Program (Grant No.  JPMJFR2237); International Joint Research
Promotion Program of Osaka University; RCNP COREnet; the Yamada
Science Foundation; the Murata Science Foundation; the Grant for
Overseas Research by the Division of Graduate Studies (DoGS) of Kyoto
University; and the Universidad Nacional Aut\'onoma de M\'exico -
DGAPA program PASPA and grant PAPIIT AG102023.

Finally, we wish to express our sorrow for the loss of Professor
Kichiji Hatanaka, and our gratitude for his guidance as collaborator
and cospokesperson of the TUCAN collaboration until his passing.
%\end{acknowledgments}

\bibliography{initial.bib}

\end{document}